\documentclass[lettersize,journal]{IEEEtran}
\usepackage{amsmath,amsfonts}
\usepackage{algorithmic}
\usepackage{algorithm}
\usepackage{array}
\usepackage{multirow}
\usepackage[caption=false,font=normalsize,labelfont=sf,textfont=sf]{subfig}
\usepackage{textcomp}
\usepackage{stfloats}
\usepackage{url}
\usepackage{verbatim}
\usepackage{graphicx}
\usepackage{color} 
\usepackage{booktabs}
\usepackage{makecell}
\usepackage{cite}
\begin{document}
	
	\title{
		A Dynamic Equivalent Method for PMSG Based Wind Farms Under Asymmetrical Faults}
	\author{Dongsheng Li,~\IEEEmembership{Student Member,~IEEE}, Chen Shen,~\IEEEmembership{Senior Member,~IEEE}, 
		\thanks{This work is supported by the National Natural Science Foundation of China under Grant U2166601. (Corresponding author: Chen Shen.)
			
			Dongsheng Li and Chen Shen are with the State Key Laboratory of Power Systems, Department of Electrical Engineering, Tsinghua University, Beijing 100084, China (e-mail: lids19@mails.tsinghua.edu.cn; shenchen@mail.tsinghua.edu.cn).
	}}

\markboth{Journal of \LaTeX\ Class Files,~Vol.~14, No.~8, August~2021}%
{Shell \MakeLowercase{\textit{et al.}}: A Sample Article Using IEEEtran.cls for IEEE Journals}


\maketitle

\begin{abstract}
In this paper, a three-machine equivalent method applicable to asymmetrical faults is proposed considering the operating wind speed and fault severity. Firstly, direct-driven permanent magnet synchronous generator wind turbines (PMSGs) are clustered based on their different active power response characteristics considering the wind speed, the fault severity, and the negative sequence control strategy. Further, single-machine equivalent methods are proposed for each cluster of PMSGs. In particular, for the PMSGs with ramp recovery characteristics, a single-machine equivalent model with multi-segmented slope recovery is proposed, which can more accurately reflect the characteristics of the wind farm during the fault recovery. Moreover, an iterative simulation method is proposed to obtain the required clustering indicators before the actual occurrence of faults. Therefore, the proposed equivalent method can be used to analyze any anticipated contingency. Eventually, the effectiveness of the proposed method is verified on a modified IEEE 39-bus system.

\end{abstract}


\section{Introduction}
\IEEEPARstart{D}{ue} to the advantages of renewability and environmental friendliness, wind power has been widely developed. However, because of the randomness and fluctuation of wind power, it can have a significant impact on the stability of the power system \cite{wang2019approaches}. Therefore, it becomes an important issue to study the impact of large-scale wind power on power systems. However, a wind farm typically contains tens or hundreds of wind turbines. If every wind turbine is modeled in detail, it will cause problems such as large memory consumption and low simulation efficiency \cite{zou_survey_2014}. Hence, in order to analyze wind farms more efficiently, it is urgent to propose an accurate dynamic equivalent method for wind farms.

Nowadays, most studies focus on the equivalent modeling of wind farms under symmetrical faults. However, most of the faults in power systems are asymmetrical faults \cite{anderson1995analysis, grainger1999power}, which will cause overvoltage on the non-faulted phase \cite{goksu2013impact, camacho2012flexible}. In order to study the response characteristics of a wind farm under asymmetrical faults more efficiently, it is necessary to develop a dynamic equivalent method for wind farms under asymmetrical faults. In addition, when studying the effectiveness of different negative sequence control strategies at the wind farm level \cite{neumann2015enhanced}, equivalent modeling of the wind farm is of great importance to improve the simulation efficiency. As a result, equivalent models for wind farms are required to improve the efficiency of wind farm analysis under asymmetrical faults.

There are two types of equivalent methods for wind farms: single-machine equivalent methods and multi-machine equivalent methods. The first method usually utilizes indicators such as wind speed \cite{brochu_validation_2011} and pitch angle \cite{jin2018equivalent} to equalize the wind farm into a single wind turbine. However, these methods ignore the differences in response characteristics among wind turbines, making it difficult to accurately model the wind farm. Multi-machine equivalent methods divide wind turbines into several groups and establish a single-machine equivalent model for each group. For example, some equivalent methods utilize wind speeds \cite{li_practical_2018,wang2020wind} or geographical location \cite{ali2012wind} to cluster wind turbines. However, these methods do not consider the impact of fault severity on wind turbine response characteristics. When the operating wind speed and geographical location are unchanged, the response characteristics of wind turbines can belong to different groups depending on the severity of faults \cite{li2022Study}. It is inaccurate to cluster wind turbines without considering the fault conditions.

In \cite{ding2016equivalent, GAO2015Consider}, wind turbines are clustered into different groups according to the operation characteristics of crowbars. However, the wind turbine response characteristics are not only related to the crowbar action characteristic but also to the control strategy and the fault severity degree. It is not accurate to cluster wind turbines by crowbar action characteristics alone. In addition, the crowbar action characteristic in asymmetrical faults also depends on the negative sequence voltage and the negative sequence control strategy, making these methods challenging to apply to asymmetrical faults. There are also methods that adopt variables such as wind speeds, terminal voltages, terminal currents \cite{zou2015fuzzy,chen2012dynamic}, and rotational speeds \cite{fang2017application} as clustering indicators. Then, clustering algorithms are applied to cluster the wind turbines. However, these indicators are all positive sequence indicators under symmetrical faults. Due to the negative sequence control, these positive sequence indicators are not sufficient to reflect the operating characteristic differences among wind turbines when asymmetrical faults occur. Therefore, the current wind farm equivalent methods are not applicable to asymmetrical faults.

There are also equivalent methods that consider adaptability in asymmetrical faults. In \cite{xue2023dynamic}, wind turbines are divided into groups based on the terminal voltage obtained from the power flow calculation. Further, an equivalent method for the zero-sequence network of the wind farm is proposed to improve the effectiveness of the equivalent model in asymmetrical faults. However, this method improves the effectiveness of the equivalent method under the asymmetrical fault only by estimating the zero-sequence impedance of the collection network. Without considering the impact of the negative-sequence component and negative-sequence control strategy of the wind turbine, the method cannot cluster wind turbines accurately. In \cite{liu2019dynamic}, an equivalent method based on the density peak clustering algorithm was proposed. The authors hold a positive view on whether the model can accommodate asymmetrical faults. However, the article does not analyze asymmetrical faults, and the clustering indicators do not consider the negative sequence component and negative sequence control strategy. The applicability of the method is still in doubt under asymmetrical faults. In \cite{Wang2019Modeling}, the negative-sequence control strategy to mitigate power and DC-link voltage oscillations is considered. An equivalent method suitable for symmetrical and asymmetrical faults is established using an improved BP algorithm. However, only the effectiveness of the method in solving short-circuit currents is proven. The output active power and reactive power of the wind farm before and after equivalence are not analyzed. These studies attempt to propose equivalent methods applicable to asymmetrical faults, but all have limitations. Moreover, there is no literature that presents the theoretical correlation between the severity of the fault and the response characteristics of wind turbines under asymmetrical faults, making it hard to cluster wind turbines accurately.

After the fault clears, wind turbines typically limit the recovery rate of the active current to reduce mechanical stress. In this case, the active power of the wind turbine will recover at a certain slope. The durations of the active power recovery of wind turbines operating at different wind speeds and voltage drops are different. In the detailed model, the recovery process of the wind farm will be composed of multiple different slopes. It is necessary to consider the differences among wind turbines during the recovery process. In \cite{chao2021analytical,li2017improved}, the active power reference value of each doubly fed wind turbine during the fault recovery period is calculated through analytical calculation, and the sum is used as the active power reference value of the equivalent wind turbine. However, in direct-driven permanent magnet synchronous generator wind turbines (PMSGs), the active power reference value is determined by the constant DC-link voltage control, making it difficult to analytically solve the active power reference value of each wind turbine. As a result, this method cannot be directly applied to PMSGs.

The response characteristics of wind turbines under asymmetrical faults are related to their negative sequence control strategy. Currently, there are two commonly used negative sequence control strategies, one for mitigating active power and DC-link voltage oscillations \cite{alepuz2009control,hu2016instantaneous} and the other for balancing the grid voltage by reducing the negative sequence voltage \cite{neumann2015enhanced,goksu2013impact}. In this paper, considering the converter capacity limitation, the above two negative sequence control strategies are theoretically analyzed separately to find the correspondence between external fault conditions and wind turbine response characteristics based on PMSGs. In the equivalent modeling process, the differences in the recovery characteristics of each PMSG are considered. Moreover, in order to obtain the clustering indicators before the occurrence of the fault, a static equivalent model of the PMSG in the positive and negative sequence network is constructed, and the clustering indicators of each wind turbine are calculated by an iterative simulation method, making the method applicable to anticipated contingencies. The main contributions are listed as follows:
\begin{itemize}
\item[1)] A clustering method considering wind speeds, fault severities, and negative sequence control strategies is proposed for PSMGs. All possible response characteristics under two different negative control strategies are analyzed. Further, the clustering boundaries are derived theoretically.
\item[2)] Single-machine equivalent models are introduced for each subgroup of PMSGs. For the cluster of PMSGs with ramp recovery characteristics, a multi-segment slope recovery equivalent model is proposed to reflect the differences of each PMSG during the fault recovery period.
\item[3)] An iterative simulation method for solving the clustering indicators is presented, which can analyze anticipated contingencies. The actual PCC voltage and clustering indicators can be obtained before the fault actually occurs.
\end{itemize}

The rest of the paper is organized as follows: the structure and control strategy of the PMSG is presented in Section II. A clustering method is put forward in Section III. Different single-machine equivalent models are proposed for different clusters of PMSGs in Section IV. An iterative simulation method for solving the clustering indicators is introduced in Section V. The proposed equivalent method is verified in a modified IEEE 39-bus system in Section VI. Conclusions are drawn in Section VII.

\section{Structure and Control Strategy of PMSGs}
An asymmetrical fault can cause imbalance in the grid voltage, and due to the existence of negative sequence active power, the DC-link capacitor voltage of PMSG also exhibits twice fundamental frequency fluctuations. For PMSGs under asymmetrical faults, there are two commonly used negative-sequence control strategies. The first one aims to balance the grid-side voltage, and the other one aims to mitigate twice fundamental frequency oscillations in the DC-link. This section will introduce the structure of the PMSG used in this paper and the implementation methods of these two negative sequence control strategies. 

\subsection{Structure of PMSGs}
The machine-side converter of the PMSG consists of a diode rectifier circuit and a boost converter circuit. The grid-side converter consists of a controlled inverter bridge composed of insulated gate bipolar transistors (IGBT). In addition, a chopper circuit is connected in parallel with the DC-link capacitor. The models of other parts of the PMSG, such as the wind turbine, the drive train system, and the synchronous generator, are consistent with common modeling methods, which can be found in \cite{wang2021adaptive}. 

As for the control system, the machine-side converter of the PMSG can control the terminal current of the synchronous generator according to the boost circuit, thereby changing the rotor speed to ensure that the wind turbine operates in the optimal tip speed ratio state, thus achieving the maximum power point tracking (MPPT) control. On the DC side, when the voltage of the DC capacitor reaches the threshold, the chopper circuit will operate to prevent overvoltage. The control strategy of the grid-side converter will be introduced in detail in the following parts. The structure of the PMSG is shown in Fig. \ref{fig_1}.

\subsection{Negative Sequence Control Strategy for Mitigating the Grid Voltage Unbalance}
During normal operation, the grid-side converter of PMSG adopts the positive sequence voltage-oriented vector control. The $dq$-axis components of the grid voltage can be derived as:
\begin{equation}
\begin{cases}
v_{d}=e\\
v_{q}=0\label{eq1}
\end{cases}
\end{equation}
where $e$ is the magnitude of grid voltage space vector; $v_{d}$ and $v_{q}$ are the $dq$-axis components of the grid voltage, respectively.

The active and reactive output power of GSC are:
\begin{equation}
\begin{cases}
P=\frac{3}{2}ei_{d}\\
Q=\frac{3}{2}ei_{q}\label{eq2}
\end{cases}
\end{equation}
where $i_{d}$ and $i_{q}$ are the $d$-axis and $q$-axis currents of grid side, respectively.

During normal operation, the reference value of active power is obtained by the constant DC-side voltage control, while the reference value of reactive power is generally maintained at 0 to achieve the unit power factor control. When an asymmetrical fault occurs, the PMSG needs to inject an appropriate amount of positive sequence reactive current into the grid and absorb negative sequence reactive current from the grid to reduce the degree of unbalance in the grid voltage according to the grid code \cite{GB19963}. The reference values of the positive and negative sequence reactive currents in the studied PMSG are:
\begin{equation}
\begin{cases}
I_{t}^+=K^+ (0.9- U^+) I_{N},(0.6\le  U^+ \le 0.9)\\
I_{t}^-=K^- U^- I_{N}
\label{eq3}
\end{cases}
\end{equation}
where the superscript "$+$" and "$-$" denote the positive and negative sequence components, respectively; $I_{t}^+$ is the reference value of positive sequence reactive current injected by the PMSG; $I_{t}^-$ is the reference value of negative reactive current absorbed by the PMSG; $K^+$ and $K^-$ are the positive and negative sequence reactive current factors, respectively; $U^+$ and $U^-$ are the magnitudes of positive and negative sequence voltages in per unit, respectively; $I_{N}$ is the rated current of the PMSG. 


After meeting the requirement of the grid code, the remaining capacity of the converter should be used to output possible maximum positive sequence active power to maintain the stability of the DC-link capacitor voltage \cite{neumann2015enhanced}. Meanwhile, the negative sequence active current is maintained at 0. Thus, the reference value of negative sequence current and positive sequence $q$-axis current can be derived by:
\begin{equation}
\begin{cases}
&I_{ref}^-=I_t^-\\
&I_{qref}^+=I_t^+ \label{ad1}
\end{cases}
\end{equation}
where $I_{ref}^-$ and $I_{qref}^+$ are the reference value of negative sequence current and positive sequence $q$-axis current, respectively.

In order to keep the output current within the maximum current limit of the converter, the positive and negative sequence currents  should satisfy the following inequality:
\begin{equation}
\left| I_{ref}^- \right|+\left| I_{ref}^+ \right| \le I_{max} \label{ad2}
\end{equation}

Combining equations \eqref{eq3}, \eqref{ad1} and \eqref{ad2}, the reference value of positive sequence active current can be derived by:
\begin{equation}
\begin{cases}
&I_{dref}^+=\min\{I_{dref1}\text{,}I_{dmax}^+\}\\
&I_{dmax}^+=\sqrt{(I_{max}-K^-  U^- I_{N})^2-I_{t}^{+2}} \label{eq4}
\end{cases}
\end{equation}
where $I_{dref}^+$ is the reference value of the positive sequence active current; $I_{dref1}$ is obtained by the constant DC-link voltage control, which is the same as the control strategy during normal operation; $I_{dmax}^+$ is the maximum value of positive sequence active current; $I_{max}$ is the maximum current allowed through the converter.

In summary, the positive and negative sequence $dq$-axis components of the terminal current of the PMSG can be derived by \cite{neumann2015enhanced,goksu2013impact}:
\begin{equation}
\begin{cases}
&I_{dref}^+=\min\{I_{dref1}\text{,}I_{dmax}^+\}\\
&I_{qref}^+=K^+ (0.9- U^+) I_{N},(0.2\le U^+ \le 0.9)\\
&I_{dref}^-=-K^-U_{q}^-I_{N}\\
&I_{qref}^-=K^-  U_{d}^- I_{N}
\label{eq5}
\end{cases}
\end{equation}
where $I_{dref}^-$ and $I_{qref}^-$ are reference values of negative sequence $dq$-axis components of terminal currents of the PMSG, respectively; $U_{d}^-$ and $U_{q}^-$ are negative sequence $dq$-axis components of the terminal voltage in per unit, respectively.

\begin{figure*}[t]
	\centering
	\includegraphics[width=7 in]{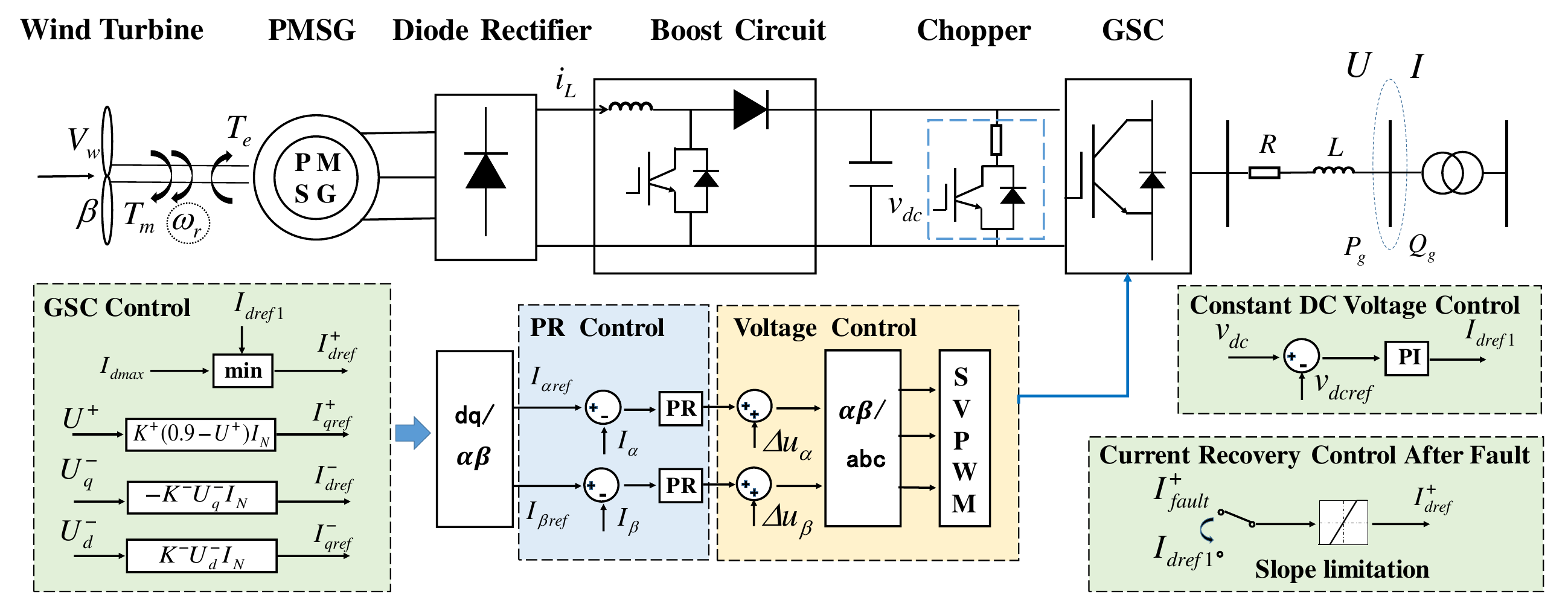}
	\caption{Topology of the PMSG.}
	\label{fig_1}
\end{figure*}

Under the control strategy, the PMSG can inject and absorb the required reactive power while outputting the positive sequence active power to maintain the stability of the DC-link capacitor voltage. If the positive sequence active current is lower than the pre-fault value due to the converter capacity limitation after the fault clears, the recovery rate of the active current is limited to reduce the mechanical stress on the PMSG \cite{feltes2009comparison,fortmann2013new}. The control strategy is also shown in Fig. \ref{fig_1}.

\subsection{Negative Sequence Control Strategy for Mitigating the Oscillation in Active Power}
When using a negative sequence control strategy for mitigating the twice fundamental frequency oscillations, the reference values of currents can be derived by:
\begin{equation}
\begin{bmatrix} I_{dref}^+ \\ 
I_{qref}^+\\
I_{dref}^-\\
I_{qref}^-
\end{bmatrix}=
\begin{bmatrix} 
U_d^+ & U_q^+ & U_d^- & U_q^- \\ 
U_q^+ & -U_d^+ & U_q^- & -U_d^- \\ 
U_d^- & U_q^- & U_d^+ & U_q^+ \\ 
U_q^- & -U_d^- & -U_q^+ & U_d^+ 
\end{bmatrix}^{-1}\times \frac{2}{3}
\begin{bmatrix} 
P_{ref} \\ 
Q_{ref} \\
0\\
0
\end{bmatrix}
\label{eq6}
\end{equation}
where $U_d^+$, $U_q^+$, $U_d^-$ and $U_q^-$ are positive and negative sequence $dq$-axis components of terminal voltages, respectively; $P_{ref}$ and $Q_{ref}$ are the DC components of the output active and reactive power, respectively.

For convenience of calculation, the reference value of reactive power is set to 0. Additionally, in order to maintain the DC-link voltage, the active power is set to the pre-fault value. If the output active power is limited by the converter capacity, the reference value of active power will be set to the maximum value within the converter capacity. At this time, the reference values for currents can be derived by:
\begin{equation}
\begin{bmatrix} I_{dref}^+ \\ 
I_{qref}^+\\
I_{dref}^-\\
I_{qref}^-
\end{bmatrix}=\frac{2P_{ref}}{3D}
\begin{bmatrix} 
U_d^+ \\ 
U_q^+ \\ 
-U_d^- \\ 
-U_q^- 
\end{bmatrix}
\label{eq7}
\end{equation}
where $P_{ref}$ and $D$ can be expressed as:
\begin{equation}
P_{ref}=\min\{P_{0}\text{,}P_{max}\}
\label{eq8}
\end{equation}
\begin{equation}
D=(U_d^{+})^2+(U_q^{+})^2-(U_d^{-})^2-(U_q^{-})^2
\label{eq9}
\end{equation}
where $P_{0}$ is the pre-fault active power of the PMSG; $P_{max}$ is the maximum active power under converter capacity limitation, which will be introduced in the following section.

\section{Clustering Boundaries for Active Power Dynamic Response Characteristics}

The active power dynamic response characteristics of PMSGs can be divided into two parts: the response characteristics during the fault and the response characteristics after the fault clears. In this section, the active power dynamic response characteristics of a single PMSG will be classified based on different wind speeds and terminal voltages, considering the negative sequence control strategy.

\subsection{Clustering Boundaries Under the Negative Sequence Control for Mitigating Unbalanced Voltage}
If the converter capacity is sufficient, the output active power of the PMSG can restore to the pre-fault value during the fault. If the external fault is severe, most of the converter's capacity is used to mitigate the unbalanced voltage. The output active power of the PMSG is limited by the converter's capacity and unable to reach the pre-fault value. Therefore, the active power characteristics of a PMSG can be divided into two groups during the fault: 1) the active power can restore to the pre-fault value; 2) the active power is limited by the converter's capacity.

Since the energy accumulated by the double-frequency component of active power on the DC-link capacitor in one cycle is almost 0, it can be considered that the DC-link capacitor voltage is only related to the DC component of active power. When the output DC active power during the fault is equal to the pre-fault value and the current of the converter is at the maximum value at the same time, the pre-fault power is the critical active power of these two groups of response characteristics under the fault. Due to the fast response of PMSGs, it can be assumed that PMSGs are able to adjust the outputs $dq$-axis currents to the reference value immediately \cite{guo2021data}. Therefore, the relationship between the critical initial active power and the voltages is as follows:
\begin{equation}
\begin{cases}
&P_0=P_{fault}=P_{cri1} \\
&I_{dref}^+=I_{dmax}^+ \\
&P_{fault}=\frac{3}{2}(I_{dref}^+U_d^++I_{qref}^+U_q^++I_{dref}^-U_d^-+I_{qref}^-U_q^-)
\label{eq10}
\end{cases}
\end{equation}
where $P_0$ is the pre-fault active power of the PMSG; $P_{fault}$ is the DC component of the output active power of the PMSG during the fault; $P_{cri1}$ is the first critical active power. Due to the positive sequence voltage-oriented vector control, $U_q^+$ equals to 0. Substituting \eqref{eq7} into \eqref{eq10}, the first critical active power can be derived as:
\begin{equation}
\begin{aligned}
&I_{dmax}^+=\sqrt{(I_{max}-K^-U^- I_{N})^2-(K^+ (0.9-U^+) I_{N})^2} \\
&P_{cri1}=\frac{3}{2}I_{dmax}^+U_d^+ 
\label{eq11}
\end{aligned}
\end{equation}

After the fault clears, the positive sequence voltage amplitude is close to 1. The active power depends on the positive sequence $d$-axis current according to \eqref{eq4}. If the positive sequence $d$-axis current at the moment of fault clear ($I_d^{f}$) is higher than the pre-fault $d$-axis current ($I_{d0}$), the output active power will rise above the pre-fault value and return to the pre-fault value after a short period of oscillation after the fault clears. If $I_d^f<I_{d0}$, the active power will recover at a fixed rate due to the limitation on the recovery rate of the active current. At this time, the second critical initial active power can be derived by:
\begin{equation}
\begin{cases}
&I_{dmax}^+=I_{dcri2} \\
&P_{cri2}=\frac{3}{2}I_{dcri2}U_{d0}^+
\label{eq12}
\end{cases}
\end{equation}
where $I_{dcri2}$ is the second critical initial $d$-axis current; $P_{cri2}$ is the second critical active power; $U_{d0}^+$ is the positive sequence $d$-axis voltage during the normal operation, which is close to 1.

Based on the above analysis, we can classify the active power response characteristics into the following three categories:
\begin{itemize}
	\item[1)] When $P_0<P_{cri1}$, we can get $I_{dref}^{+}=I_{dref1}$. The DC component of the active power of PMSG has recovered to the pre-fault value during the fault, which can keep the DC-link voltage at the reference value. 
	\item[2)] When $P_{cri1}\le P_0<P_{cri2}$, we can get $I_{dref}^{+}=I_{dmax}^+$. The DC component of the active power of PMSG is lower than the pre-fault value while the positive sequence active current is higher than the pre-fault active current during the fault. Thus, when the voltage restores, the output active power will rise above the pre-fault value and return to the pre-fault value after a short period of oscillation.
	\item[3)] When $ P_0 \ge P_{cri2}$, we can get  $I_{dref}^{+}=I_{dmax}^+$. There is a ramp recovery process after the fault clears.
\end{itemize}

\begin{figure}[!htb]
	\centering
	\includegraphics[width=3 in]{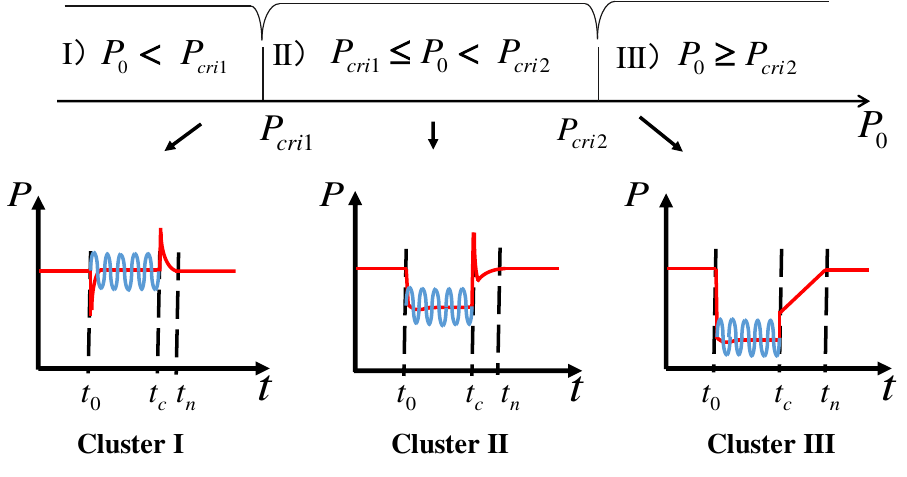}
	\caption{Wind power curve of PMSG-WTG.}
	\label{fig_2}
\end{figure}

The three types of active response characteristics are shown schematically in Fig. \ref{fig_2}. The blue curve represents the active power response characteristics, while the red curve shows the DC component of the active power. In the figure, $t_0$ denotes the time of fault start, $t_c$  denotes the time of fault clear, $t_n$ denotes the time when the PMSG returns to normal.

\begin{figure}[!htb]
	\centering
	\includegraphics[width=2.5 in]{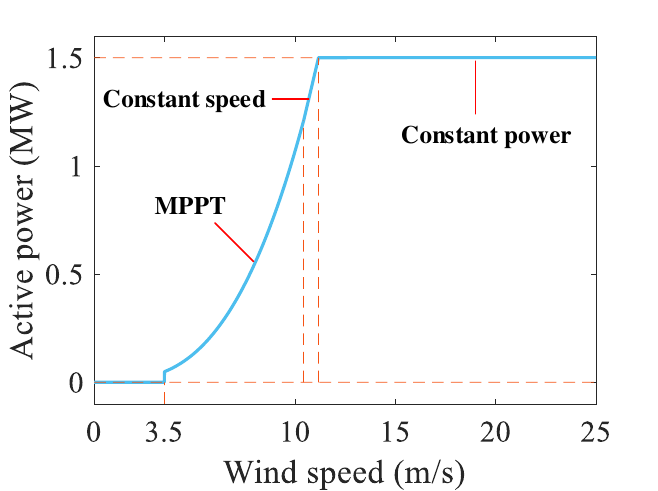}
	\caption{Wind power curve of PMSG-WTG.}
	\label{fig_3}
\end{figure}

Moreover, the critical wind speeds of each sub-cluster can be calculated from the critical power and the wind power curve of the PMSG. The critical wind speeds can be derived by:
\begin{equation}
\begin{aligned}
V_{cri1}=f^{-1}(P_{cri1})\\
V_{cri2}=f^{-1}(P_{cri2})\label{eq13}
\end{aligned}
\end{equation}
where $f^{-1}$ is the inverse function of the wind power curve. The wind power curve is shown in Fig. \ref{fig_3}.

\begin{figure}[!htb]
	\centering
	\includegraphics[width=3 in]{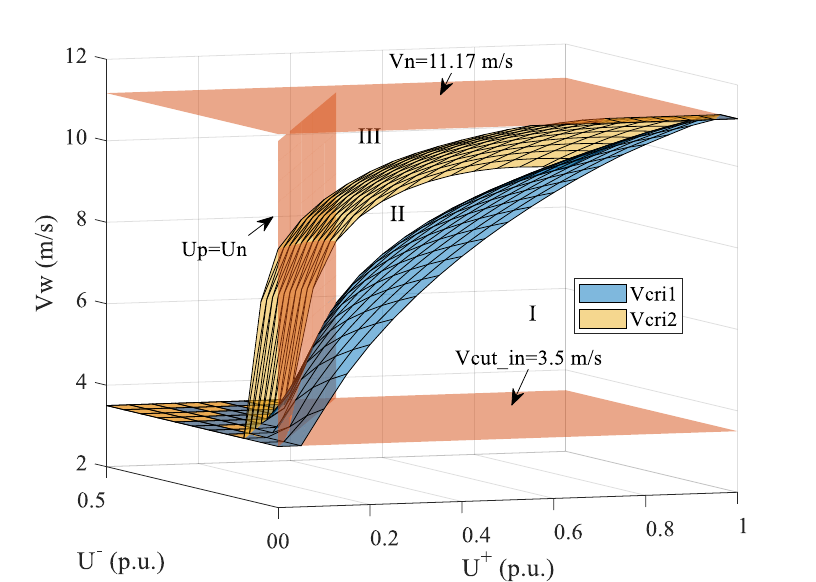}
	\caption{Classification boundaries of response characteristics.}
	\label{fig_4}
\end{figure}

According to \eqref{eq11} and \eqref{eq12}, the clustering boundary conditions of the three types of response characteristics are shown in Fig. \ref{fig_4}. The two red planes above and below refer to the rated wind speed and the cut-in wind speed of the PMSG, respectively. When the PMSG operates at or above its rated wind speed, the output power is the rated active power due to the existence of pitch angle control. The red plane perpendicular to the $U^+$$U^-$ plane represents the situation where the positive and negative sequence voltage drops are equal. The positive sequence voltage is usually greater than the negative sequence voltage in an asymmetrical fault, so only the right-hand part of the plane needs to be analyzed. The three clusters labeled in Fig. \ref{fig_4} correspond to the three clusters in Fig. \ref{fig_2}. 

\begin{figure}[!htb]
	\centering
	\includegraphics[width=3 in]{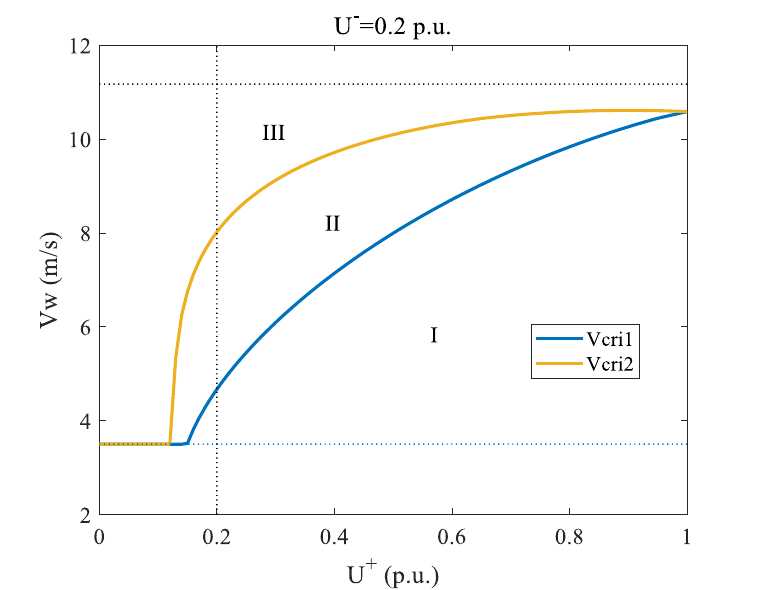}
	\caption{Clustering boundaries when $U^-=0.2 p.u.$.}
	\label{fig_5}
\end{figure}

In order to show the clustering boundary more clearly, we take the case where the negative sequence voltage drops to 0.2 p.u. as an example and draw the clustering boundaries based on the positive sequence voltage and the wind speed as shown in Fig. \ref{fig_5}. When the positive and negative sequence terminal voltages and the operating wind speed of a PMSG are known, it can be quickly determined which cluster the active power response characteristics of the PMSG belong to by using the clustering boundaries shown in Fig. \ref{fig_4}.

\subsection{Clustering Boundaries Under the Negative Sequence Control for Eliminating Oscillations in active power}
Under the negative sequence control for eliminating the twice fundamental frequency oscillations in active power, the response characteristics of the PMSG can also be divided into three categories, as shown in Fig. \ref{fig_2}. By comparing the maximum output active power during the fault with the initial active power, it can be determined whether the active power characteristics belong to Cluster I or Cluster II in Fig. \ref{fig_2}. When the PMSG output the maximum active power, the positive and negative sequence currents should satisfy the following equation:
\begin{equation}
\sqrt{(I_{dref}^+)^2+(I_{qref}^+)^2} +\sqrt{(I_{dref}^-)^2+(I_{qref}^-)^2}=I_{max}
\label{eq14}
\end{equation}

Substituting \eqref{eq7} into \eqref{eq14}, we can get:
\begin{equation}
\frac{2P_{max}}{3D}( U^+ +U^-)=I_{max}
\label{eq15}
\end{equation}

Simplify \eqref{eq15} and $P_{cri1}$ can be derived by:
\begin{equation}
\begin{cases}
&P_{cri1}=P_{max} \\
&P_{max}=\frac{3}{2}(U^+ -U^- )I_{max}
\label{eq16}
\end{cases}
\end{equation}

Similarly, comparing the magnitude of the positive sequence $d$-axis current at the moment of fault clear with the pre-fault $d$-axis current can determine whether there is a ramp recovery process in the PMSG. If there is a ramp recovery process, the positive sequence active current is lower than the pre-fault value and the active power must not rise to the pre-fault value during the fault. As a result, the reference active power is equal to the maximum active power during the fault ($P_{ref}=P_{max}$) according to \eqref{eq8}. The second critical initial active power can be derived by:
\begin{equation}
\begin{cases}
&I_{dcri2}=I_{dref}^+ \\
&I_{dref}^+=\frac{2P_{max}}{3D}U_d^+
\label{eq17}
\end{cases}
\end{equation}
where $I_{dcri2}$ is the second critical initial $d$-axis current.

Substituting \eqref{eq16} into \eqref{eq17}, we can get:
\begin{equation}
\begin{cases}
&I_{dcri2}=\frac{U_d^+I_{max}}{ U^+ +U^-} \\
&P_{cri2}=\frac{3}{2}I_{dcri2}U_{d0}^+
\label{eq18}
\end{cases}
\end{equation}

\begin{figure}[!htb]
	\centering
	\includegraphics[width=3 in]{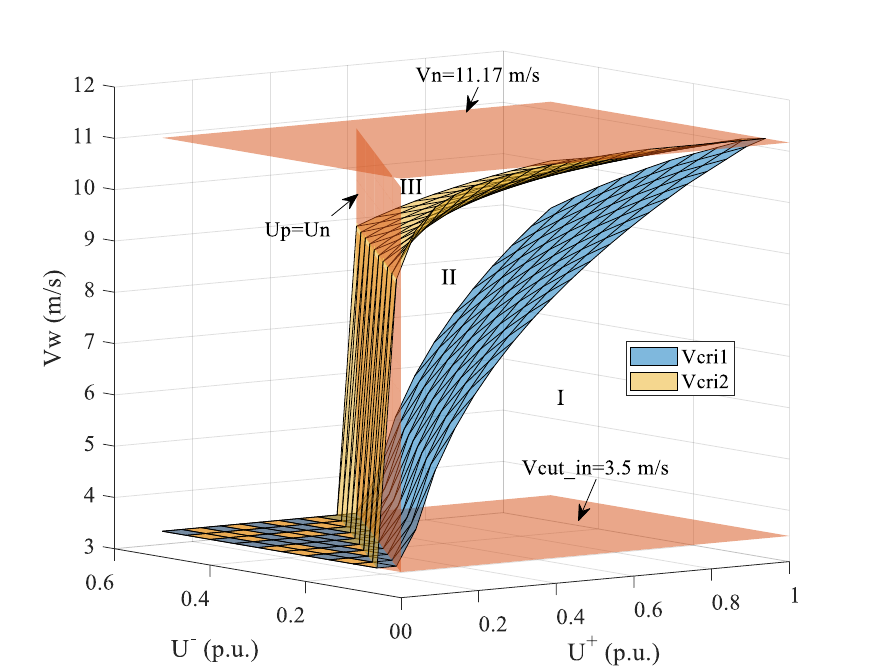}
	\caption{Classification boundaries of response characteristics under the second negative sequence control.}
	\label{fig_6}
\end{figure}

The critical wind speeds considering the negative sequence control strategy mentioned above are shown in Fig. \ref{fig_6}. The critical wind speeds are limited between the cut-in wind speed and the rated wind speed.

\begin{figure}[!htb]
	\centering
	\includegraphics[width=2.5 in]{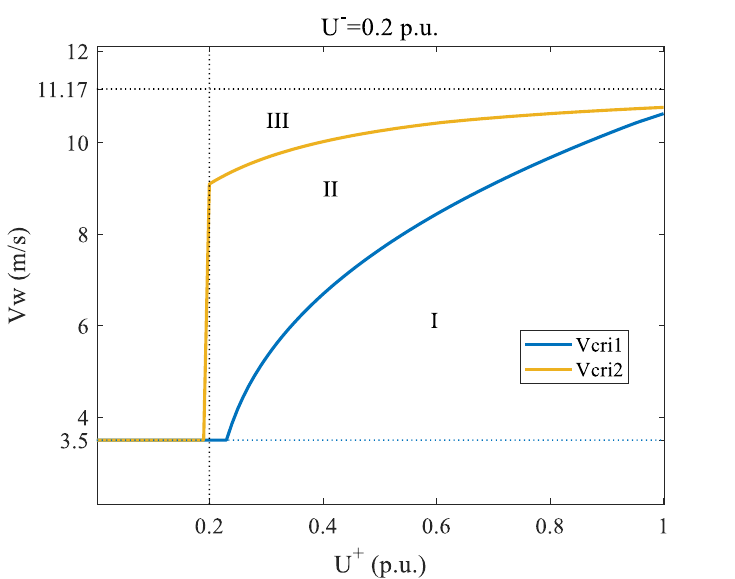}
	\caption{Clustering boundaries when $U^-=0.2 p.u.$. under the second negative sequence control strategy.}
	\label{fig_7}
\end{figure}

Similarly, the critical wind speeds are shown in Fig. \ref{fig_7} when the negative sequence voltage drops to 0.2 p.u.

\section{Dynamic Equivalent Modeling Method for PMSGs}
After clustering the PMSGs, a single-machine equivalent models can be adopted for each cluster of PMSGs. This section will introduce the dynamic equivalent models for the above three clusters of PMSGs.
\subsection{Capacity Weighted Equivalent Method}
The capacity weighted equivalent method is always employed to aggregate the wind turbines in the same cluster \cite{zou2015fuzzy,wang2021adaptive}. For the PMSGs in the first and second clusters shown in Fig. \ref{fig_2}, the characteristics of active power response are still the same as that of a single PMSG after accumulation. Therefore, the capacity weighted equivalent method can be used to equalize these two clusters of PMSGs. The effectiveness of this method has been theoretically analyzed in \cite{li2017improved}. The equivalent parameters can be calculated by:
\begin{equation}
\begin{cases}
S_{eq}=\sum_{i=1}^{N}S_i\\[2mm]
X_{eq}=\frac{X}{N},R_{eq}=\frac{R}{N}\\[2mm]
H_{t\_eq}=\frac{1}{N}\sum_{i=1}^{N}H_{t\_i},H_{g\_eq}=\frac{1}{N}\sum_{i=1}^{N}H_{g\_i}
\label{eq19}
\end{cases}
\end{equation}
where $N$ is the number of PMSGs in one cluster; $S$ is the capacity of a PMSG; $X$ and $R$ are the reactance and resistance of stator; $H_t$ is the turbine inertia time constant; $H_g$ is the generator inertia time constant; the subscript ``$eq$'' denotes the equivalent value; the subscript ``$i$" denotes the parameter of the $i$th PMSG.

\subsection{Equivalent Model with Multi-Segment Slope Recovery}
\begin{figure}[!htb]
	\centering
	\includegraphics[width=3 in]{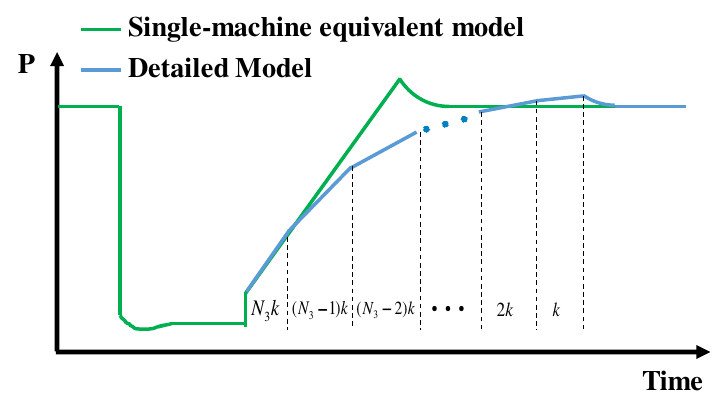}
	\caption{Differences between the single-machine equivalent model and the detailed model.}
	\label{fig_9}
\end{figure}

For PMSGs in Cluster III, since the terminal voltage and the operating wind speed of each PMSG are different, the duration of the slope recovery process is also different. The sum of the active power of PMSGs in Cluster III will recover at different rates in different time periods, as shown by the blue line in Fig. \ref{fig_9}. If the capacity weighted equivalent method is still used for this group of PMSGs, the active power of the equivalent PMSG will recover to the pre-fault value at a fixed slope, as shown by the green line in Fig. \ref{fig_9}. In order to present an accurate equivalence of the fault recovery process, an equivalent model with multi-segment slope recovery is proposed.

Firstly, we can calculate the duration of the slope process of each PMSG by their terminal voltages and operating wind speeds. Then, based on the recovery duration of each PMSG, we can limit the recovery rate of the equivalent PMSG at different values in different time periods. In this way, an accurate equivalence of the fault recovery process is achieved. 

For the PMSGs in Cluster III, the positive sequence $d$-axis current of each wind turbine is limited by the capacity of the converter according to \eqref{eq12} and \eqref{eq18}. Thus, the positive sequence $d$-axis currents at the moment of fault clear of each PMSG are $I_{dcri2}$. Taking the negative sequence control for mitigating unbalanced voltage as an example, the duration of the slope process of each PMSG can be derived by:
\begin{equation}
t_i=(I_{d0i}-I_{dcri2\_i})/k
\label{eq20}
\end{equation}
where $t_i$ is the time duration of the slope recovery process of the $i$th PMSG in Cluster III; $I_{d0i}$ is the pre-fault $d$-axis current of the $i$th PMSG; $I_{dcri2\_i}$ is the second critical $d$-axis current of the $i$th PMSG; $k$ is the maximum recovery rate of the $d$-axis current.

The pre-fault active power and pre-fault $d$-axis current of the $i$th PMSG in Cluster III can be derived as follows:
\begin{equation}
\begin{aligned}
P_{0i}=f(V_{wi}) \\
I_{d0i}=\frac{2P_{0i}}{3U_{d0}^+} \label{eq21}
\end{aligned}
\end{equation}
where $V_{wi}$ is the operating wind speed of the $i$th PMSG.

Substituting \eqref{eq11}, \eqref{eq12} and \eqref{eq21} into \eqref{eq20}, the time duration of the $i$th PMSG can be derived as follows:
\begin{equation}
\begin{aligned}
&t_i=(\frac{2f(V_{wi})}{3e}-I_{dcri2\_i})/k\\
&I_{dcri2\_i}=\sqrt{(I_{max}-K^- U_i^-I_{N})^2-(K^+ (0.9-U^+_i) I_{N})^2}
\end{aligned}
\label{eq22}
\end{equation}
	
\begin{figure}[!htb]
	\centering
	\includegraphics[width=3 in]{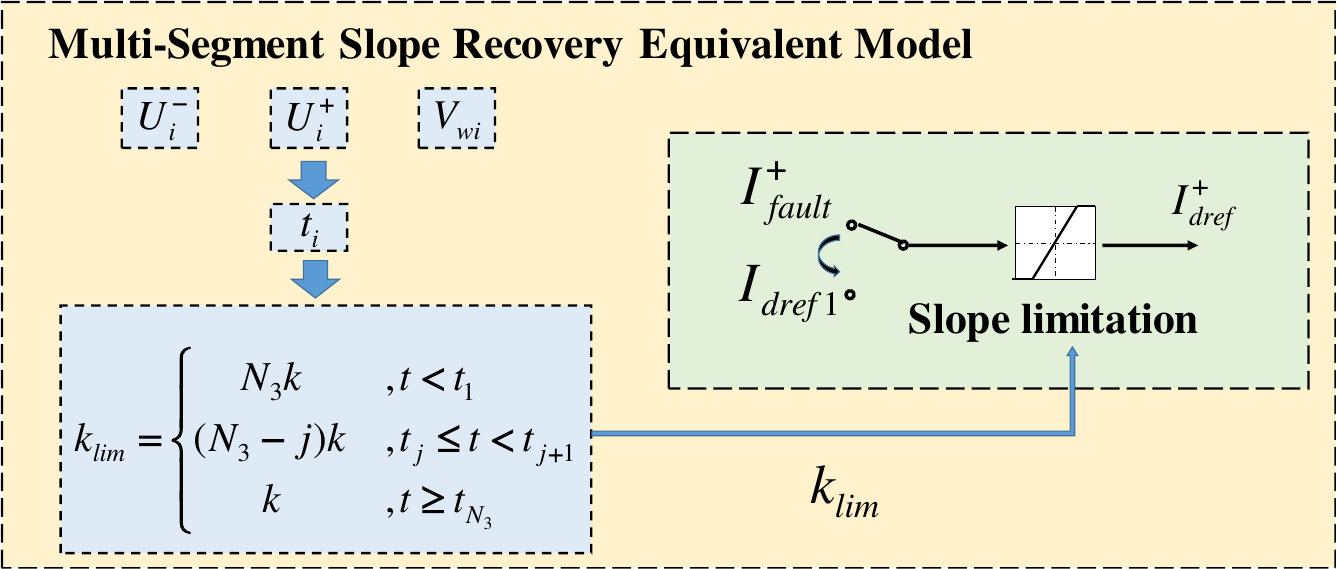}
	\caption{Multi-segment slope recovery equivalent method.}
	\label{fig_10}
\end{figure}

We can see that $t_i$ is determined by the PMSG parameters, the terminal voltage, and the operating wind speed of the $i$th PMSG. Sorting $t_i$ from the smallest to the largest, the $d$-axis current recovery rate of the equivalent PMSG after the fault clears is limited by:
\begin{equation}
k_{lim}=
\begin{cases}
\ \ \ \  N_3k & ,  t <t_1\\
(N_3-j)k & , t_j\le t<t_{j+1},j=1\cdots N_3-1\\
\ \ \ \ \ \ k & , t\ge t_{N_3}
\label{eq23}
\end{cases}
\end{equation}
where $N_3$ is the number of PMSGs in Cluster III; $t$ is timed from the moment of fault clear. The recovery rate is maintained at $k$ when $t\ge t_{N_3}$ to model the overshoot process of the constant DC-link voltage control. The schematic diagram of the proposed method is shown in Fig. \ref{fig_10}.

\subsection{Equivalent Method for Collector Lines}
The equal voltage drop method is adapted to calculate the parameters of the equivalent collector lines \cite{garcia2008modelling}. Since the positive and negative sequence impedance parameters of the collector lines inside a wind farm are generally equal, we only use the equal positive sequence voltage drop method to equate the collector lines under asymmetrical faults. The average positive sequence terminal voltage of PMSGs in the same cluster can be derived by:
\begin{equation}
\dot U_{ave}^+=\frac{1}{N}\sum_{i=1}^{N} \dot U_{i}^+
\end{equation}
where $\dot U_{ave}^+$ is the average voltage phasor of PMSGs; $N$ is the number of PMSGs in the same cluster; $\dot U_{i}^+$ is the positive sequence terminal voltage phasor of the $i$th PMSG.

Since the average voltage drop before and after equivalence is equal, the parameter of the equivalent collector line can be derived by:
\begin{equation}
Z_{eq}=\frac{\dot U_{pcc}^+-\dot U_{ave}^+}{\sum_{i=1}^{N} \dot I_{i}^+}
\end{equation}
where $Z_{eq}$ is the parameter of the equivalent collector line; $\dot U_{pcc}^+$ is the positive sequence PCC voltage phasor; $\dot I_{i}^+$ is the output positive sequence current phasor of the $i$th PMSG.

The proposed method in this section requires parameters such as the terminal voltage and the output current of each PMSG, which will be obtained in the following section.

\section{Method for Solving the Clustering Indicators}
If wind speeds and terminal voltages of PMSGs are known, PMSGs can be divided into three clusters, as shown in Fig. \ref{fig_2} according to the clustering boundaries. Then, PMSGs in each sub-cluster can be modeled by a single-machine equivalent method to obtain the three-machine equivalent model of the wind farm. The wind speed of each PMSG can be obtained by prediction or measurement. However, the terminal voltage of each PMSG is not only related to the wind speed but also to the severity of the fault and the response characteristics of other components in the power system, which is difficult to obtain analytically. In this section, a positive and negative sequence terminal voltage calculation method is presented, assuming that the PCC voltage is known. Then, based on this method, the actual PCC voltage is obtained by an iterative simulation method. Eventually, the equivalent model of the wind farm can be obtained.

\subsection{Calculation Method for Terminal Voltages}
\begin{figure}[!htb]
	\centering
	\includegraphics[width=3 in]{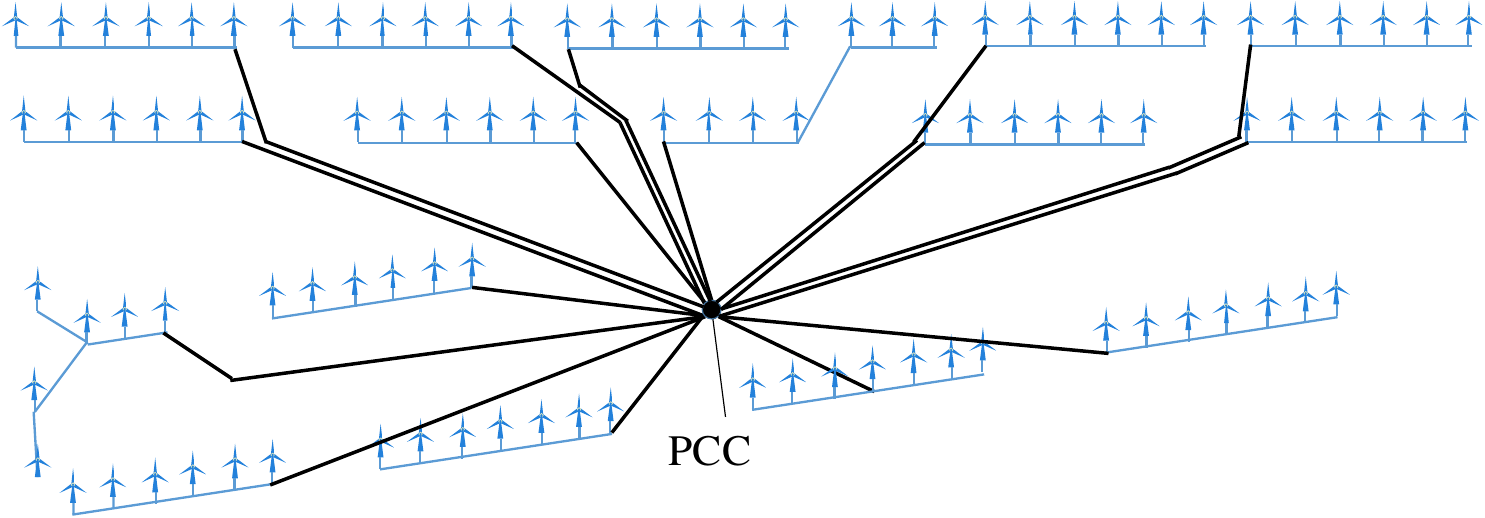}
	\caption{Wind farm topology.}
	\label{fig_11}
\end{figure}

A terminal voltage calculation method is proposed based on a real wind farm topology, as shown in Fig. \ref{fig_11}. Due to the fast response of PMSGs, the terminal voltages can be calculated by the static model of PMSGs when PCC voltage is known. In this section, we analyze the negative sequence control for mitigating unbalanced voltages as an example.

 We decouple the positive and negative sequence networks to calculate the positive and negative sequence terminal voltages of PMSGs, respectively. According to \eqref{eq5}, the output $dq$-axis currents of the PMSG in the negative sequence network are only related to the negative sequence voltage dips, while the $d$-axis current in the positive sequence network is affected by both positive and negative sequence voltages. Therefore, we can calculate the negative sequence terminal voltage under the negative sequence network first and then further calculate the positive sequence terminal voltage.
 
\subsubsection{Calculation Method for Negative Sequence Terminal Voltages of PMSGs}

In the negative-sequence network, the output active and reactive power of each PMSG can be derived by:
\begin{equation}
\begin{cases}
P_{i}^-=1.5( U_{di}^-  I_{drefi}^-+ U_{qi}^-  I_{qrefi}^-)\\
Q_{i}^-=1.5( U_{qi}^-  I_{drefi}^-- U_{di}^-  I_{qrefi}^-)
\end{cases}
\label{eq26}
\end{equation}
where $P_{i}^-$ and $Q_{i}^-$ are the output active and negative power of the $i$th PMSG in the negative-sequence network. Substituting \eqref{eq5} into \eqref{eq26}, we can get:
\begin{equation}
\begin{cases}
P_{i}^-=0\\
Q_{i}^-=-1.5K^-( U_{qi}^{-2}  + U_{di}^{-2} )I_N=-1.5K^-U_{i}^{-2}I_N
\end{cases}
\label{eq27}
\end{equation}

The current injected by the $i$th PMSG can be derived by:
\begin{equation}
\dot I_{ni}^-=(\frac{P_{i}^-+j{Q_i^-}}{\dot U_i^-})^*
\label{eq28}
\end{equation}
where $\dot I_{ni}^-$ is the injected current phasor of the $i$th PMSG in the negative sequence network; $\dot U_{i}^-$ is the negative sequence terminal voltage phasor of the $i$th PMSG.

Further, to establish the node-branch incidence matrix for each feeder of the wind farm, the branch current column vector of a feeder can be derived as:
\begin{equation}
\dot I_{b}^-=C^T\dot I_{n}^-
\label{eq29}
\end{equation}
where $C$ is the node-branch incidence matrix of the feeder line; $\dot I_{n}^-$ is the column vector of negative sequence current consisting of $\dot I_{ni}^-$; $\dot I_{b}^-$ is the negative sequence branch current column vector. Further, the node voltage drop can be derived by:
\begin{equation}
\begin{aligned}
& \Delta \dot U_b^-=Z \dot I_b^- \\
& \Delta \dot U_n^-=C\Delta \dot U_b^-
\end{aligned}
\label{eq30}
\end{equation}
where $\Delta \dot U_b^-$ and $\Delta \dot U_n^-$ are the branch voltage drop and node voltage drop, respectively; $Z$ is the branch impedance matrix of the feeder. Only the values of the diagonal elements of $Z$ are the branch impedance. Other non-diagonal elements are 0. Moreover, the voltage of each node on the feeder can be updated to:
\begin{equation}
\dot U^{-'}=\dot U_{pcc}^-+\Delta \dot U_{n}^-
\label{eq31}
\end{equation}
where $\dot U_{pcc}^-$ is the negative sequence voltage of PCC; $\dot U^{-'}$ is the column vector of the updated voltage of each node on the feeder.

Substituting \eqref{eq29} and \eqref{eq30} into \eqref{eq31}, we can get:
\begin{equation}
\dot U^{-'}=\dot U_{pcc}^-+CZC^T\dot I_{n}^-
\label{eq32}
\end{equation}

When the negative sequence voltage at PCC is known, the negative sequence terminal voltage of each PMSG can be obtained by the following steps:
\begin{itemize}
	\item[1)] Get the branch impedance matrix and the node-branch incidence matrix of each feeder based on the wind farm topology data and set all negative sequence terminal voltages of PMSGs to the PCC negative sequence voltage ($U_i^-=U_{pcc}^-$).
	\item[2)] Calculate the revised negative sequence terminal voltage of each PMSG according to \eqref{eq27}, \eqref{eq28} and \eqref{eq32}.
	\item[3)] If $\lvert \dot U^{'-}-\dot U^- \rvert<\sigma_1$, the vector of negative sequence terminal voltages of all PMSGs are available in $U^{'-}$. If not, assign the value of $\dot U^{'-}$ to $\dot U^-$ and turn to step 2. Where $\sigma_1$ is the allowable error of terminal voltages.
\end{itemize}

\subsubsection{Calculation Method for Positive Sequence Terminal Voltages of PMSGs}
	In the positive-sequence network, the output active and reactive power of each PMSG can be derived by:
\begin{equation}
\begin{cases}
&P_{0i}^+=1.5( U_{di}^+  I_{drefi}^++ U_{qi}^+  I_{qrefi}^+)\\
&Q_{0i}^+=1.5( U_{qi}^+  I_{drefi}^+- U_{di}^+  I_{qrefi}^+)
\end{cases}
\label{eq33}
\end{equation}
while the reference values of the current are rewritten here as:
\begin{equation}
\begin{cases}
&I_{qrefi}^+=K^+ (0.9- U_i^+) I_{N}\\
&I_{drefi}^+=\min\{I_{dref1i}\text{,}I_{dmaxi}\}\\
&I_{dmaxi}=\sqrt{(I_{max}-K^- \lvert U_i^-\rvert I_{N})^2-I_{qrefi}^{+2}}
\label{eq34}
\end{cases}
\end{equation}
where $I_{dref1i}$ is the $d$-axis current reference value of the constant DC-link capacitor voltage control. In order to maintain the DC voltage, the following equation should be satisfied:
\begin{equation}
U_{d0}^+I_{d0i}=U_i^+I_{dref1i}
\label{eq35}
\end{equation}
while $U_{d0}^+$ is close to 1, $I_{dref1i}$ can be derived by:

\begin{equation}
I_{dref1i}=\frac{I_{d0i}}{U_i^+}
\label{eq36}
\end{equation}

Since $U_i^-$ has been solved in the previous section, it can be seen from \eqref{eq33}, \eqref{eq34} and \eqref{eq36} that the output active and reactive power of PMSGs in the positive sequence network are only related to the operating wind speeds and the positive sequence terminal voltages. According to \eqref{eq28}-\eqref{eq32}, we can also solve the positive sequence terminal voltage of each PMSG using the same method proposed in Section V.A(1). The only difference is that we should replace the negative sequence components with the corresponding positive sequence components.

As a result, if the PCC voltage is known, we can solve the terminal voltage of each PMSG using the proposed method in Section V.A.

\subsection{Iterative Simulation Method for Solving the PCC Voltage}

When analyzing the anticipated contingencies, we need to consider the effect of fault severity on the response characteristics of PMSGs. However, PCC voltage is difficult to obtain before a fault actually occurs. Therefore, an iterative simulation method to solve the PCC voltage is presented in this section. The method can obtain the PCC voltage before the occurrence of the fault, avoiding the simulation of the detailed model to obtain the clustering indicators. Thus, the proposed equivalent method is applicable to the analysis of anticipated contingencies. The specific steps are shown as follows:

\begin{itemize}
	\item[1)] Input the wind speed of each PMSG and set $U_{pcc}^+=1$,$U_{pcc}^-=0$.
	\item[2)] Let $\dot U_{pcc}^+=U_{pcc}^+\angle 0 ^\circ$ and $\dot U_{pcc}^-=U_{pcc}^-\angle 0 ^\circ$. Calculate the positive and negative sequence terminal voltages of each PMSG-WTG using the method proposed in Section V.A.
	\item[3)] Build the equivalent model of the wind farm using the method proposed in Section III and Section IV.
	\item[4)] Simulation analysis of the anticipated contingency is carried out using the established equivalent model in step 3). Moreover, the positive and negative sequence PCC voltage at the moment of the fault clear ($U_{pcc}^{+'}$ and $U_{pcc}^{-'}$) can be obtained by the result of the simulation.
	\item[5)] If $\lvert U_{pcc}^{+'}-U_{pcc}^+ \rvert<\sigma_2$ and $\lvert U_{pcc}^{-'}-U_{pcc}^- \rvert<\sigma_2$, turn to step 6. If not, set $U_{pcc}^{+}=U_{pcc}^{+'}$, $U_{pcc}^{-}=U_{pcc}^{-'}$ and turn to step 2. $\sigma_2$ is the allowable error of the PCC voltage.
	\item[6)] Since the actual PCC voltage is obtained, clustering indicators of each PMSG can be calculated by the method presented in Section V.A. Eventually, the equivalent model of the wind farm can be obtained by the method proposed in Section III and Section IV.
\end{itemize}

\section{Method Verification}
\begin{figure}[!htb]
	\centering
	\includegraphics[width=3 in]{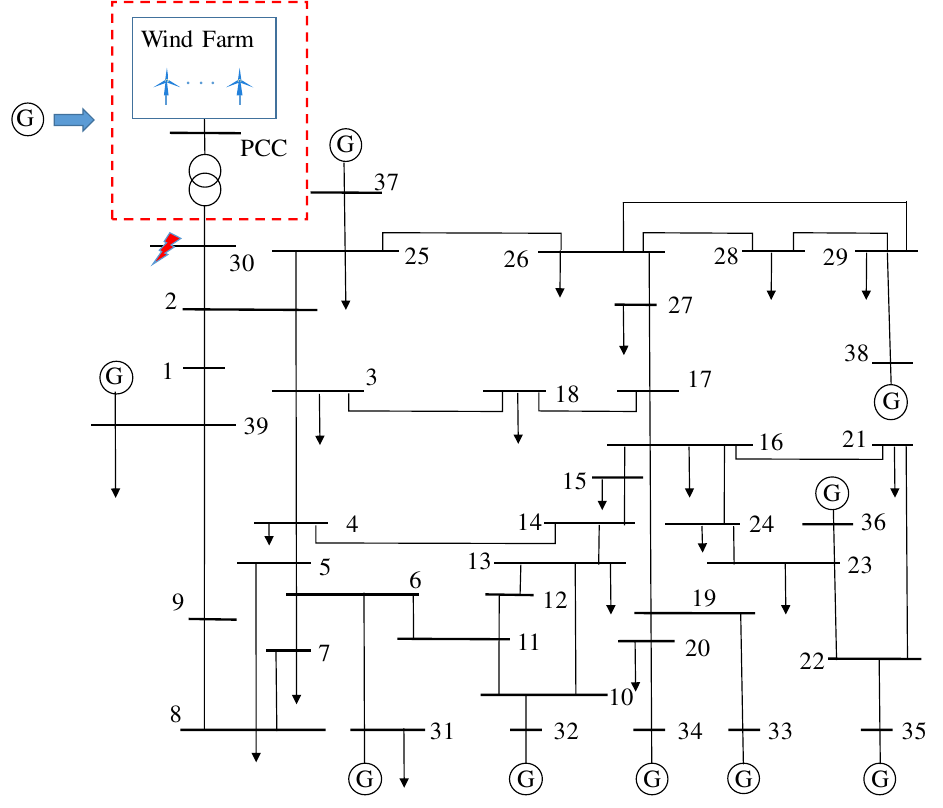}
	\caption{A modified IEEE 39-bus system.}
	\label{fig_12}
\end{figure}

\begin{figure}[!htb]
	\centering
	\includegraphics[width=3 in]{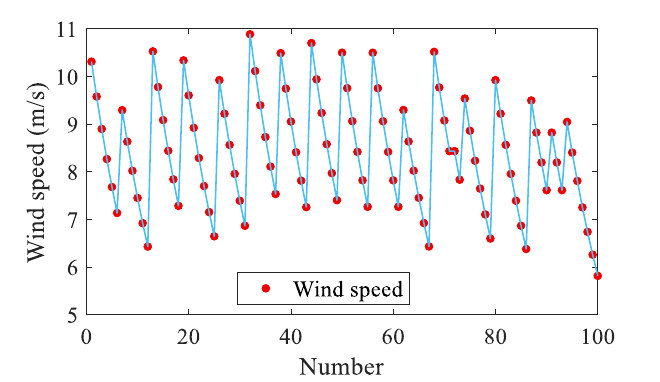}
	\caption{Wind speed of each PMSG.}
	\label{fig_13}
\end{figure}

The proposed method is verified under different faults on the CloudPSS platform \cite{song2020cloudpss, RDF_W3}. A wind farm including 100 PMSGs with a rated capacity of 1.5 MW is studied as shown in Fig. \ref{fig_11}. The wind farm is connected to node 30 of the IEEE 39-bus system through a transformer, as shown in Fig. \ref{fig_12}. The wind speeds of PMSGs are modeled with the Jensen model, assuming that wind speeds among different feeders do not affect each other. The wind speeds of PMSGs are shown in Fig. \ref{fig_13}.

\subsection{Case I: Two-phase Short Circuit Fault at Terminal of the Wind Farm}
\begin{figure}[!htb]
	\centering
	\includegraphics[width=3.5 in]{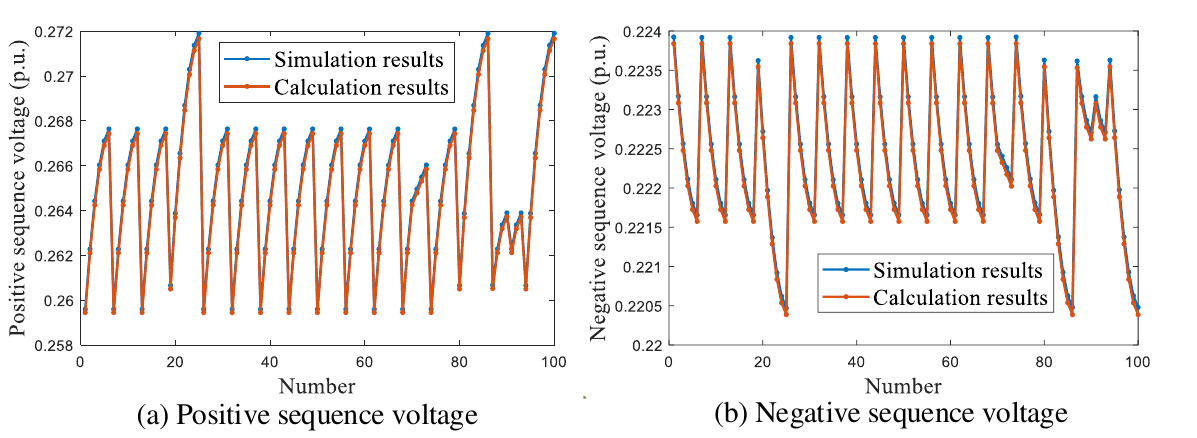}
	\caption{Terminal voltage of each PMSG.}
	\label{fig_14}
\end{figure}

The two-phase short-circuit fault between phase B and phase C starts at 3.0 s and clears at 3.2 s at node 30. Phase B and C are connected to the ground through an impedance. The PCC voltage can be obtained by the iterative simulation method proposed in Section V.B. The iterative process is shown in Table. \ref{Voltage_iter}. The PCC voltage converges after one iteration of the equivalent model. Moreover, the terminal voltage of each PMSG at the moment before the fault clearance can be calculated using the method presented in Section V.A. The calculated positive and negative sequence terminal voltage results are compared with the simulation results of the detailed model, as shown in Fig. \ref{fig_14}. The maximum percentage error in the voltage of each node is 0.089\%, which proves the correctness of the voltage calculation method proposed in Section V.A.

\begin{table}
	\renewcommand\arraystretch{1.5}
	\begin{center}
		\caption{PCC Voltage in Each Iteration of Case I.}
		\label{Voltage_iter}
		
		\begin{tabular}{ c  c  c }
			\toprule 
			Model  &  \multicolumn{2}{c}{Simulation results (p.u.)}\\
					&	 $U^+_{pcc}$  &	$U^-_{pcc}$ \\
			
			\midrule
			\makecell[c]{Initial equivalent model\\ $U^+_{pcc}=1$, $U^-_{pcc}=0$} & 0.253 & 0.226\\
			\midrule
			\makecell[c]{Equivalent model 1 \\$U^+_{pcc}=0.253$, $U^-_{pcc}=0.226$}& 0.253 & 0.226\\ 
			\midrule
			Detailed model& 0.253 & 0.226\\
			\bottomrule			
		\end{tabular}
		
	\end{center}
\end{table}

\begin{figure}[!htb]
	\centering
	\includegraphics[width=3.5 in]{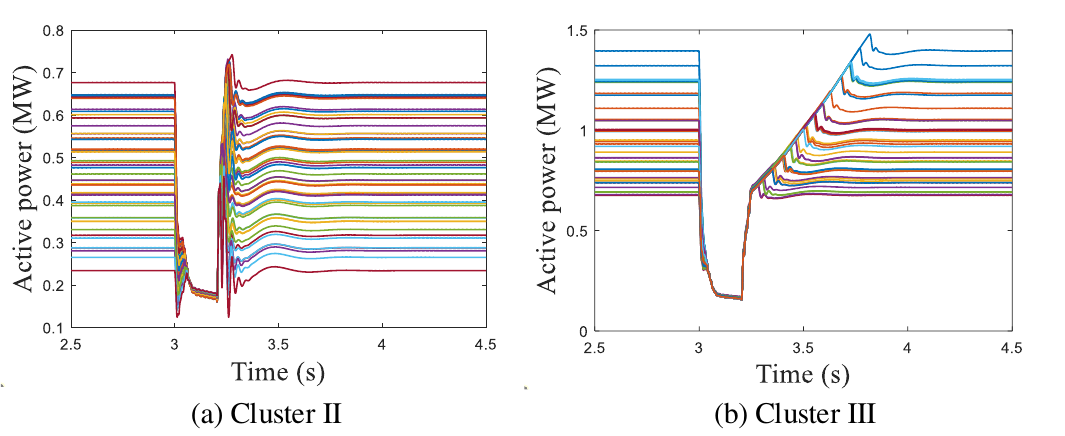}
	\caption{Active power of individual PMSG in the same cluster in Case I.}
	\label{fig_15}
\end{figure}

Further, PMSGs are clustered into three clusters based on the method proposed in Section III. The clustering result is shown in Table. \ref{Clustering Results of Case I.}. The DC components of individual PMSG active power in different clusters are shown in Fig. \ref{fig_15}. The characteristics are consistent with the analytical results, which verifies the correctness of the proposed method in Section III.

\begin{table}
	\renewcommand\arraystretch{1.2}
	\begin{center}
		\caption{Clustering Results of Case I.}
		\label{Clustering Results of Case I.}
		
		\begin{tabular}{ c  c }
			\toprule
			Cluster & Number of PMSG-WTGs \\
			\midrule 
			Cluster 1&--\\
			\midrule
						&4, 5, 6, 9, 10, 11, 12, 16, 17, 18, 22, 23, 24,\\
			Cluster 2	&25, 29, 30,31, 36, 37, 41, 42, 43, 48, 49, 53,\\
			&54, 55, 59, 60, 61,64, 65, 66, 67, 71, 72, 73,\\
			&76, 77, 78, 79, 82, 83, 84, 85,86, 89, 90, 92,\\
			&93, 95, 96, 97, 98, 99, 100\\

			\midrule
									&1, 2, 3, 7, 8, 13, 14, 15, 19, 20, 21, 26, 27,\\ 
			Cluster 3	&28, 32, 33, 34, 35, 38, 39, 40, 44, 45, 46, 47,\\
			&50, 51, 52, 56, 57, 58, 62, 63, 68, 69, 70, 74,\\ 
			&75, 80, 81, 87, 88, 91, 94\\

			\bottomrule			
		\end{tabular}
		
	\end{center}
\end{table}

\begin{figure}[!htb]
	\centering
	\includegraphics[width=3.5 in]{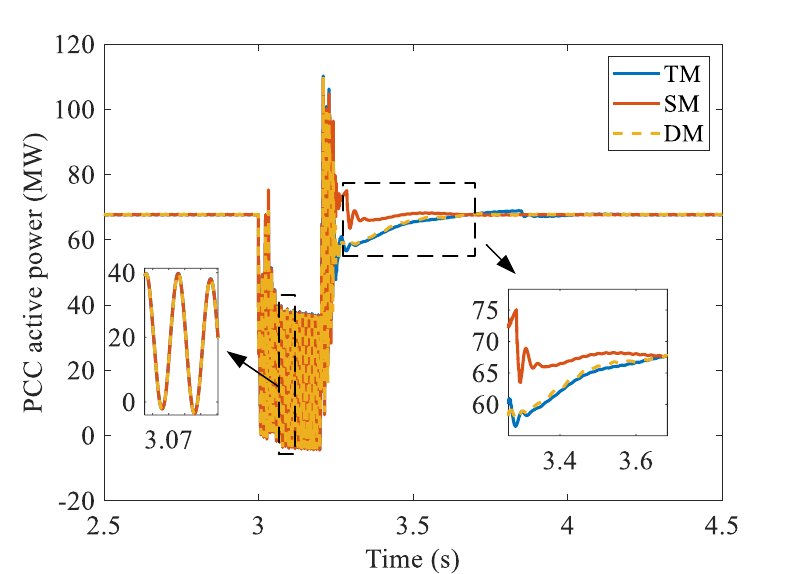}
	\caption{The comparison of active power.}
	\label{fig_16}
\end{figure}

\begin{figure}[!htb]
	\centering
	\includegraphics[width=3.5 in]{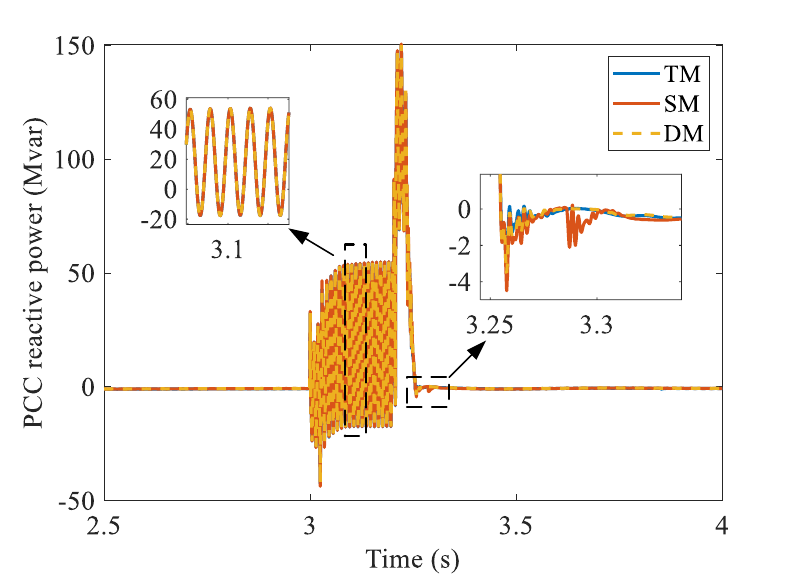}
	\caption{The comparison of reactive power.}
	\label{fig_17}
\end{figure}

In order to prove the efficiency of the dynamic equivalent method, the active and reactive power of the detailed model (DM), the single-machine equivalent model (SM) \cite{mercado2015aggregated}, and the proposed three-machine equivalent model (TM) are compared, as shown in Fig. \ref{fig_16} and Fig. \ref{fig_17}.

Due to the severe fault, the PMSGs belong to Cluster II and Cluster III in Case I. Thus, the active power response of each model is basically the same during the fault. After the fault clears, the active power of PMSGs in Cluster III will recover at a certain rate, and the TM is able to represent this partial characteristic more accurately than the SM. The simulation times and the equivalent root mean square errors (RMSEs) are shown in Table. \ref{Method_Compare}.

\subsection{Case II: One-phase Short Circuit Fault at Terminal of the Wind Farm}
In Case II, an A-phase short circuit fault starts at 3.0 s and clears at 3.2 s at node 30. Phase A is connected to the ground through a larger impedance, and the voltage dip is slighter than that in Case I. The wind speeds of PMSGs still adopt the wind speeds shown in Fig. \ref{fig_13}. Similar to Case I, the PCC voltage and the terminal voltage of each PMSG can be obtained by the iterative simulation method proposed in Section V. The PCC voltage results are $U^+_{PCC}=0.646$ and $U^-_{PCC}=0.360$. Based on the wind speed and terminal voltage of each PMSG, the clustering result can be obtained by the method presented in Section III, as shown in Table. \ref{Clustering Results of Case II.}.

\begin{table}
	\renewcommand\arraystretch{1.2}
	\begin{center}
		\caption{Clustering Results of Case II.}
		\label{Clustering Results of Case II.}
		
		\begin{tabular}{ c  c }
			\toprule
			Cluster & Number of PMSG-WTGs \\
			\midrule 
       		 &4, 5, 6, 9, 10, 11, 12, 17, 18, 22, 23, 24, 25,\\
Cluster 1    &29, 30, 31, 36, 37, 42, 43, 48, 49, 54, 55, 60,\\
    	     &61, 64, 65, 66, 67, 73, 76, 77, 78, 79, 83, 84,\\
       		 &85, 86, 89, 90, 92, 93, 96, 97, 98, 99, 100\\
			\midrule
			 &3, 7, 8, 15, 16, 21, 27, 28, 34, 35, 40, 41, 46,\\
Cluster 2	 &47, 52, 53, 58, 59, 62, 63, 70, 71, 72, 74, 75,\\
			 &81, 82, 87, 88, 91, 94, 95\\
			\midrule
Cluster 3	 &1, 2, 13, 14, 19, 20, 26, 32, 33, 38, 39, 44,\\
			 &45, 50, 51, 56, 57, 68, 69, 80\\
			
			\bottomrule			
		\end{tabular}
		
	\end{center}
\end{table}

\begin{figure}[!htb]
	\centering
	\includegraphics[width=3.5 in]{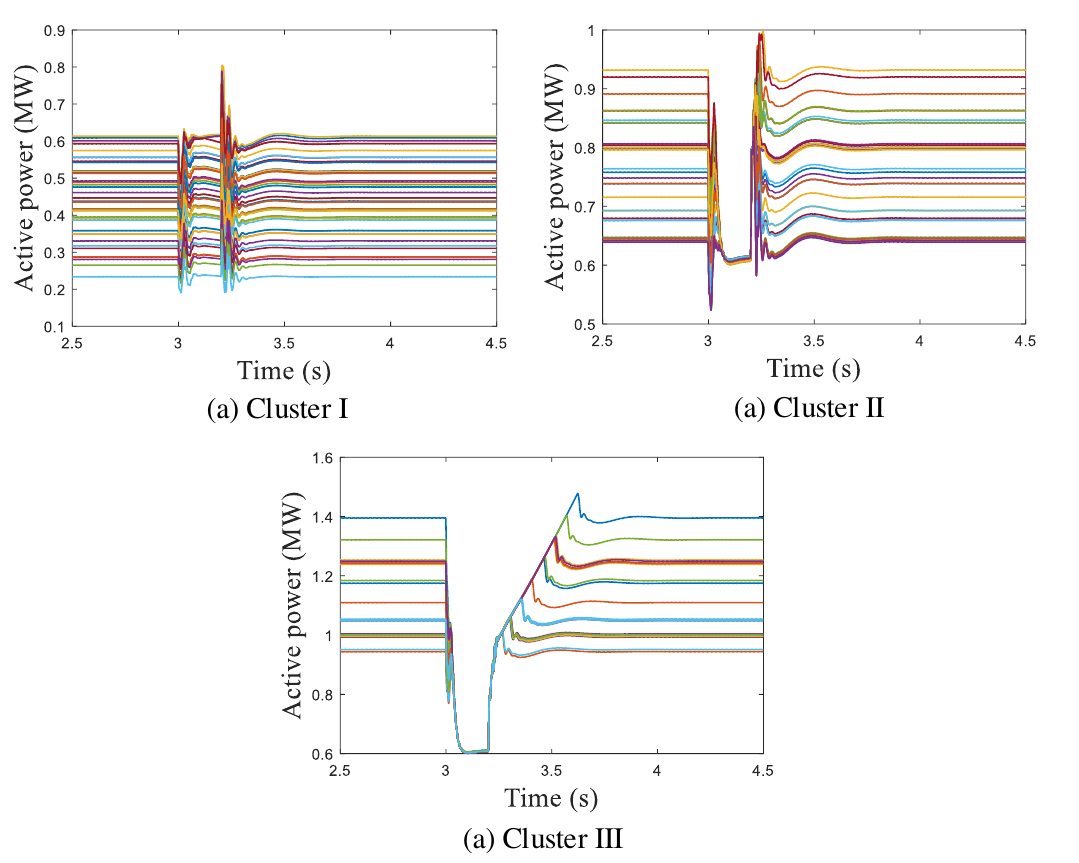}
	\caption{Active power of individual PMSG in the same Cluster in Case II.}
	\label{fig_18}
\end{figure}

\begin{figure}[!htb]
	\centering
	\includegraphics[width=3.5 in]{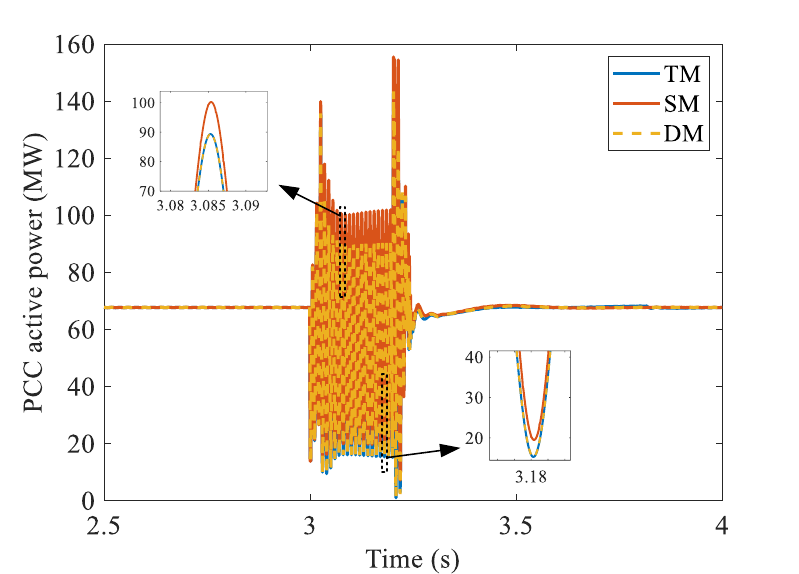}
	\caption{The comparison of active power.}
	\label{fig_19}
\end{figure}

\begin{figure}[!htb]
	\centering
	\includegraphics[width=3.5 in]{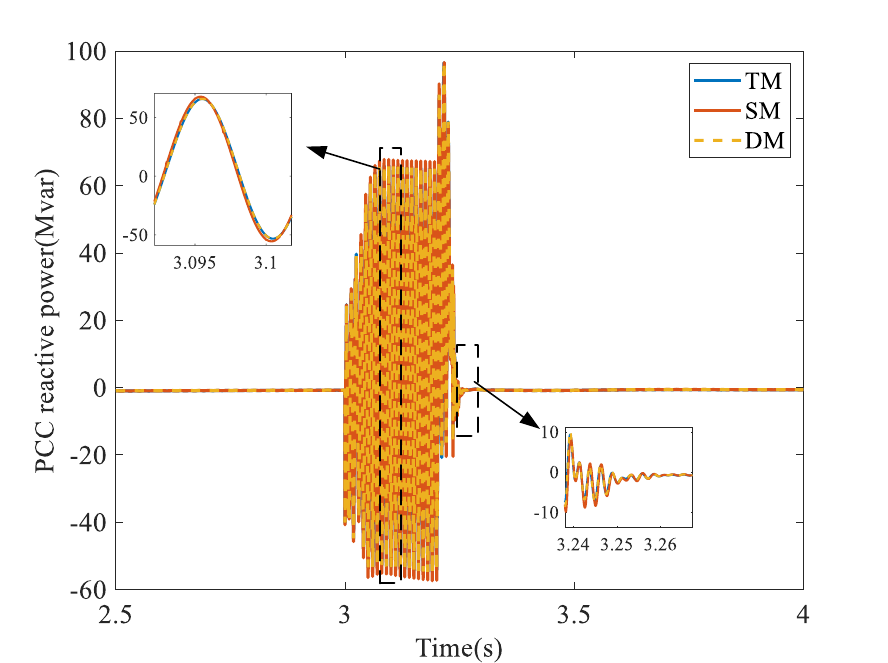}
	\caption{The comparison of reactive power.}
	\label{fig_20}
\end{figure}

The DC components of individual PMSG active power in different clusters are shown in Fig. \ref{fig_18}. The active and reactive powers of different models are compared in Fig. \ref{fig_19} and Fig. \ref{fig_20}.

In Case II, most of the PMSGs belong to Cluster I and there is only a small number of PMSGs have a ramp recovery process. Therefore, the TM and SM are able to perform the active power characteristics accurately after the fault clears. However, during the fault period, the TM can present the active power characteristics more accurately compared to the SM because the active powers of PMSGs in different clusters are limited by different factors.

\begin{figure}[!htb]
	\centering
	\includegraphics[width=3.5 in]{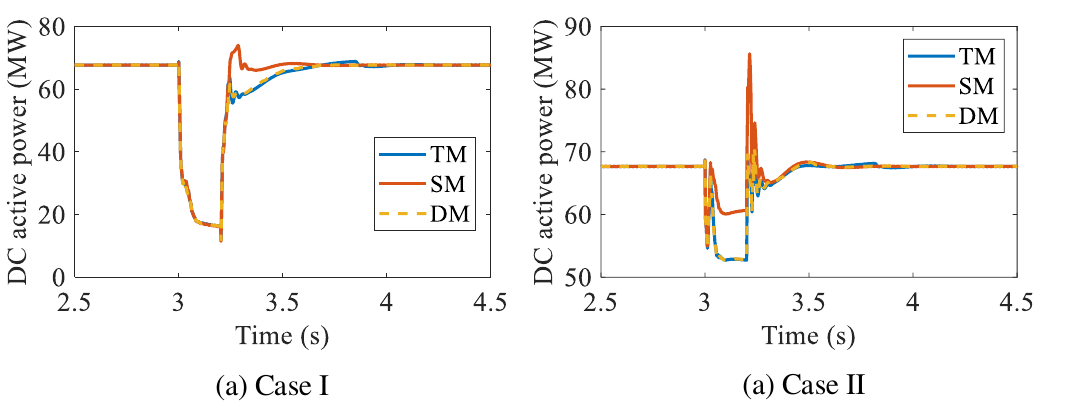}
	\caption{The comparison of active power.}
	\label{fig_21}
\end{figure}

In order to reflect the effectiveness of the proposed method more clearly, we present the DC components of the active power of different models, as shown in Fig. \ref{fig_21}. The SM is inaccurate after the fault clears in Case I and inaccurate during the fault in Case II. In summary, the traditional equivalent method cannot reflect the active power differences among PMSGs in different clusters, while the proposed method is able to achieve accurate equivalence under different faults.

\begin{table}[htbp] 
	\centering
	\caption{Simulation Times and Equivalent Errors of different methods.}
	\label{Method_Compare} 
	\begin{tabular}{ccccc} 
		\toprule 
		& & DM & TM & SM \\
		\midrule 
		\multirow{2}{*}{Case I} & \makecell[c]{Simulation\\ time (s)} & 549 & 20 & 14  \\
		& \makecell[c]{Equivalent\\ error (\%)} 					  & -- & 1.23 & 6.10 \\
		\midrule 
		\multirow{2}{*}{Case II} & \makecell[c]{Simulation\\ time (s)}& 491 & 16 & 11  \\
		& \makecell[c]{Equivalent\\ error (\%)} 				      & -- & 0.41 & 10.01  \\

		\bottomrule 
	\end{tabular} 
\end{table}

\section{Conclusion}
Under asymmetrical faults, the proposed equivalent method takes the effects of the operating wind speed, the fault severity, and the negative sequence control into consideration. The PMSGs are clustered based on their active power response characteristics. Further, the equivalent models are proposed for each cluster of PMSGs, respectively. Moreover, an iterative simulation method for calculating the clustering indicators is presented to make the proposed method applicable to the anticipated faults. Thus, the difficulty in obtaining clustering indicators is solved. Eventually, the proposed method is validated on a modified IEEE 39-bus system, and the effectiveness of the proposed method is demonstrated by the significant reduction of simulation time compared with the detailed model and the improvement of equivalence accuracy compared with the traditional equivalent method.

\bibliographystyle{IEEEtran}
\bibliography{ref_asy}
\end{document}